%% file: main.tex
\documentclass[dvipsnames]{IEEEtran}

\IEEEoverridecommandlockouts
\usepackage{amsmath,amssymb,amsfonts}
\usepackage{xfrac}
\usepackage{graphicx}
\usepackage{xparse}
\usepackage{siunitx}
\usepackage[breaklinks,colorlinks]{hyperref}
\hypersetup{
   colorlinks=true,
   linkcolor=black,
   filecolor=magenta,      
   urlcolor=black,
}
\usepackage[capitalize]{cleveref}
\usepackage{pifont} 
\def\cone{\ding{192}} 
\def\ctwo{\ding{193}}
\def\cthree{\ding{194}}
\def\cfour{\ding{195}}
\def\cfive{\ding{196}}

\usepackage[style=ieee, backend=bibtex, defernumbers=true]{biblatex}
\input{acronyms.tex}

\newcommand{\fcut}{f_\text{cut}}

\newcommand{\Vpr}{\text{V}_\text{pr}}
\newcommand{\Vpd}{\text{V}_\text{pd}}
\newcommand{\VgprMirr}{\text{V}_\text{mirr}}
\newcommand{\VprAmp}{\text{V}_\text{prAmp}}
\newcommand{\IprMirr}{\text{I}_\text{prMirr}}

\begin{document}

\title{SciDVS: A Scientific Event Camera with 1.7\% Temporal Contrast Sensitivity at 0.7 lux%
}

\author{\IEEEauthorblockN{Rui Graca, Sheng Zhou, Brian McReynolds, Tobi Delbruck\\}
\IEEEauthorblockA{\textit{Sensors Group, Inst. of Neuroinformatics, UZH-ETH Zurich, 
Zurich, Switzerland} \\
rpgraca,shengzhou,bmac,tobi@ini.uzh.ch, \url{https://sensors.ini.ch}}
}

\maketitle


\begin{abstract}
This paper reports a Dynamic Vision Sensor (DVS) event camera that is 6x more sensitive at 14x lower illumination than existing commercial and prototype cameras. Event cameras output a sparse stream of brightness change events.  Their high dynamic range (HDR), quick response, and high temporal resolution provide key advantages for scientific applications that involve low lighting conditions and sparse visual events.
However, current DVS are hindered by low sensitivity, resulting from shot noise and pixel-to-pixel mismatch.
Commercial DVS have a minimum brightness change threshold of $>$10\%.
Sensitive prototypes achieved as low as 1\%, but required kilo-lux illumination.
Our SciDVS prototype fabricated in a 180nm CMOS image sensor process achieves 1.7\% sensitivity at chip illumination of 0.7\,lx and 18\,Hz bandwidth.
Novel features of SciDVS are (1) an auto-centering in-pixel preamplifier providing intrascene HDR and increased sensitivity, (2) improved control of bandwidth to limit shot noise, and (3) optional pixel binning, allowing the user to trade spatial resolution for sensitivity. 
\end{abstract}

\section{Introduction}
\label{sec:intro}
\begin{figure}
\centering
\includegraphics[width=1.0\columnwidth]{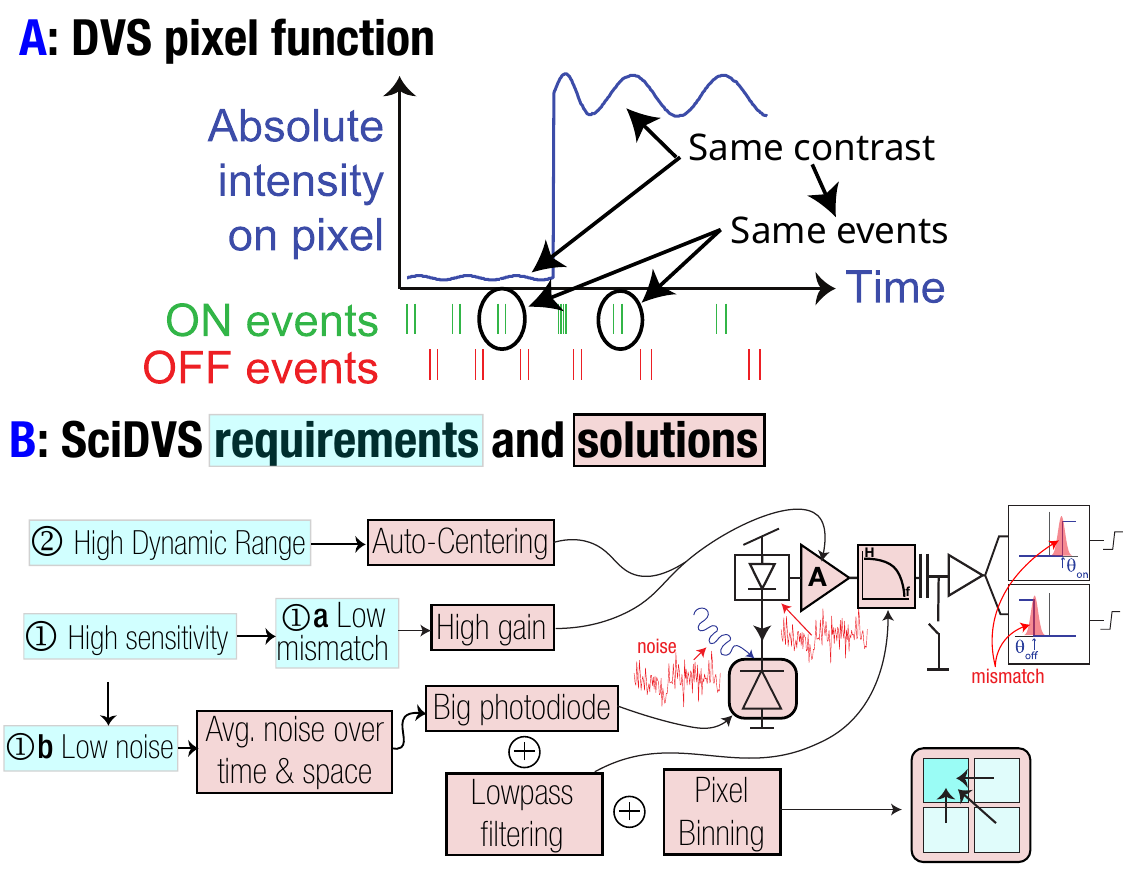}
\caption{\textbf{A}: Principle of DVS operation with varying \xx{hdr} illumination~\cite{lichtsteiner2005aerlogarithmic}.
\textbf{B}: Scientific DVS requirements and SciDVS solutions.}
\label{fig:concept}
\end{figure}

\cref{fig:concept}A illustrates how a \xx{dvs} pixel encodes brightness changes into sparse, activity-driven events~\cite{gallego2022eventbased,lichtsteiner2005aerlogarithmic,brandli2014latencyglobal,serrano2013contrastsensitivity,yang2015dynamicvision,moeys2018sensitivedynamic,taverni2017invivoimaging,taverni2018frontandback,son2017dynamicvision,finateu2020backilluminated,guo2023waferstacked,niwa2023eventbased,kodama2023rgbhybrid}.
Its logarithmic response provides \xx{hdr}. 
The low-latency and sparse events enable high-speed vision with low computational complexity.

Sensing small temporal contrast with an event camera would benefit applications such as fluorescent imaging of neural activity and space domain awareness. 
These applications require a sensitivity of $<$10\% and are characterized by sparse periods of rapid activity, require \xx{hdr}, and must operate with dim lighting. 
They currently rely on high speed sCMOS cameras with limited \xx{dr} and highly redundant frame output.
Recent commercial \xx{dvs} development focuses on reducing pixel size \cite{finateu2020backilluminated,son2017dynamicvision,niwa2023eventbased,kodama2023rgbhybrid,guo2023waferstacked}, which impacts sensitivity by increasing shot noise and pixel-to-pixel mismatch \cite{graca2023shininglight}.
Previous sensitive \xx{dvs} cameras achieved higher sensitivity in bright settings \cite{serrano2013contrastsensitivity,yang2015dynamicvision,moeys2018sensitivedynamic} by adding a preamplifier between photoreceptor and change amplifier.
The increased gain allows smaller \xx{nct} without having too many ``hot'' pixels (induced by event threshold mismatch) that have extremely high noise event rates.

However, no prior \xx{dvs} camera has demonstrated sub-10\% sensitivity under dim illumination conditions, mainly because noise in \xx{dvs} pixels was not understood. Sensitive \xx{dvs} designs required very bright illumination (on the order of \si{\kilo\lux}).
Bright illumination is not available for neural imaging, 
where the fluorescence signal emitted by the tissue is a minuscule fraction of the incident light; 
or for space applications, where stars and satellites are dim points against a dark background. 
Under dark conditions, shot noise is the dominant factor limiting sensitivity~\cite{graca2023shininglight,rose1973vision}.
The SciDVS camera reported in this paper exploits the recent understanding~\cite{graca2023optimalbiasing} that \xx{dvs} photoreceptor noise can reach a minimum of 2x photon shot noise by properly biasing the photoreceptor and its buffer.

\begin{figure*}[tb]
\centering
\includegraphics[width=\textwidth]{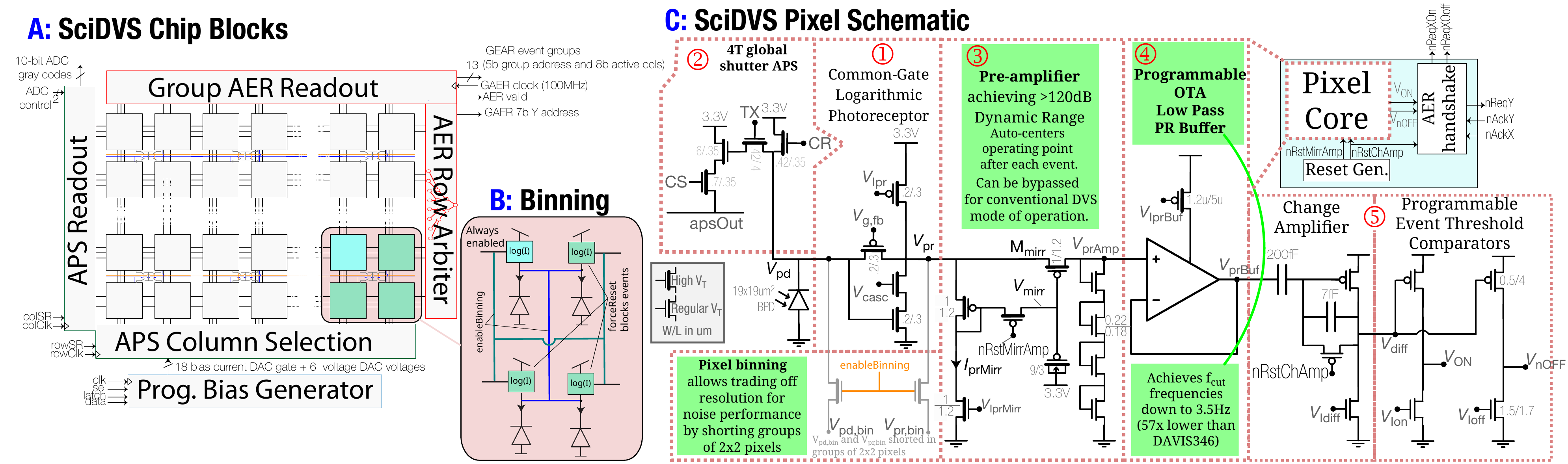}
\caption{Circuits. 
\textbf{A}: Chip block diagram
with APS readout \cite{brandli2014latencyglobal}, group AER readout \cite{son2017dynamicvision}, and bias generators \cite{yang2012addressablecurrent} \textbf{B}: \numproduct{2x2}
binning that shorts photodiode/photoreceptor outputs and can optionally disable \sfrac{3}{4} of event
generators. 
\textbf{C}: pixel schematic, with novel SciDVS features in green boxes.}
\label{fig:circuits}
\end{figure*}

Photon shot noise
introduces a fundamental limit to the visual process. As shown by Rose~\cite{rose1973vision}, 
for a single pixel to reliably detect an edge with contrast $C$, 
it needs to accumulate a number of photons proportional to $1/C^2$.
Since the minimum contrast detectable by a \xx{dvs} defines its sensitivity, when operating at the 2x shot noise limit, increasing sensitivity is equivalent to accumulating more photons. This can be done spatially, by increasing the photodiode area, or temporally, by low-pass filtering the photoreceptor output.

\cref{fig:concept}B illustrates how SciDVS addresses the requirements for a scientific event camera by combining several solutions:
\cone{}~High-sensitivity demands: \cone{}\textbf{a} low pixel-to-pixel mismatch, which is met by introducing a preamp as in \cite{serrano2013contrastsensitivity,yang2015dynamicvision,moeys2018sensitivedynamic}, and 
\cone{}\textbf{b} low noise, which is met by combining larger spatial integration using a large photodiode, as well as a novel binning technique, and longer temporal integration using an optimized low-pass filter. 
\ctwo\ The \xx{hdr} requirement is met by a novel auto-centering circuit for the preamp.

\section{Description of Key features}
\label{sec:features}


\cref{fig:circuits}A shows the SciDVS block diagram, which includes Group AER readout \cite{son2017dynamicvision}, 9b APS global shutter frames (not concurrent with events) \cite{brandli2014latencyglobal}, and programmable biases \cite{yang2012addressablecurrent}.

\cref{fig:circuits}B\&C shows the novel SciDVS binning mode. Binning is widely used for low-light CIS.
A pixel binning concept for foveated \xx{dvs} proposed photocurrent summing \cite{serrano2022electronicallyfoveated}.
SciDVS shorts both photodiodes ($\Vpd$) and photoreceptors ($\Vpr$) of \numproduct{2x2} pixel groups. Enabling \texttt{enableBinning} reduces $\text{V}_\text{rms}$($\Vpr$) by 2x. 
Due to mismatch, each pixel has a slightly different threshold, so the outputs are not necessarily redundant.
Enabling \texttt{forceReset} additionally reduces output event rate by disabling 3 of the 4 event generators in a pixel bin. 

\cref{fig:circuits}C shows the pixel circuits. 
Photocurrent generated in the buried photodiode can be steered by adjusting the biases and control signals to the \xx{dvs} common-gate photoreceptor \cone\ or to the 4T APS frame readout \ctwo{} from \cite{brandli2014latencyglobal}.
The APS readout is enabled by setting $\text{V}_\text{g,fb}=\text{V}_\text{Vdd}$  which steers photocurrent to the APS readout. The column reset \texttt{CR}, select \texttt{CS}, transfer gate \texttt{TX}, and source-follower output \texttt{apsOut} are as in \cite{brandli2014latencyglobal}. Setting \texttt{CR}=0 steers current to the \xx{dvs} photoreceptor. 

The photoreceptor output $\Vpr$ is amplified by the preamp~\cthree{}~based on the diode-stack from \cite{serrano2022electronicallyfoveated}.
The preamp mirrors the photocurrent with $M_\text{mirr}$ and then applies this current to a stack of diode-connected transistors providing a gain of $\approx7$.
The preamp output $\VprAmp$ is proportional to the log of the mirrored photocurrent, with a gain dependent only on the thermal voltage and the weak inversion slope of the transistors in the stack, hence subject to 
low mismatch.

In \cite{serrano2013contrastsensitivity,moeys2018sensitivedynamic}, the increase in gain introduced by preamplification results in a decrease in intrascene \xx{dr} to only about 60dB centered around the array-level average illumination, thus losing a key \xx{dvs} advantage.
Ref.~\cite{yang2015dynamicvision} addressed this problem by adding pixel-level adaptation of the preamp operating point using a pseudo-resistor. It provides standard \xx{dvs} \xx{hdr}, but the adaptation time after a high contrast edge is slow, temperature dependent, and mismatch-prone.

The SciDVS preamp overcomes the 
\xx{dr} limitation in~\cite{serrano2013contrastsensitivity} by introducing a 
 3T-1C pixel-level auto-centering circuit: 
After each event, \texttt{nRstMirrAmp} is temporarily activated, 
establishing a circuit path that sets $\VgprMirr$ to mirror $\IprMirr$, 
re-centering the dynamic range for that pixel.
The preamp can be bypassed (by switches not shown) for standard \xx{dvs} operation.
The preamp output $\VprAmp$ drives a 5T OTA low-pass buffer \cfour, that is designed with long, thick-gate, high $V_T$ FETs to improve the source follower stage by allowing $\fcut$ bandwidths 57x smaller than in previous \xx{dvs}, allowing the user to trade speed for sensitivity.
The change detector and threshold comparators \cfive\ use the proven topology from previous \xx{dvs}~ 
\cite{lichtsteiner2005aerlogarithmic,brandli2014latencyglobal,serrano2013contrastsensitivity,moeys2018sensitivedynamic,taverni2017invivoimaging,taverni2018frontandback,son2017dynamicvision,finateu2020backilluminated,guo2023waferstacked}
.

\section{Results and Comparison}
\label{sec:results}

\begin{figure}[tb]
\centering
\includegraphics[width=.7\columnwidth]{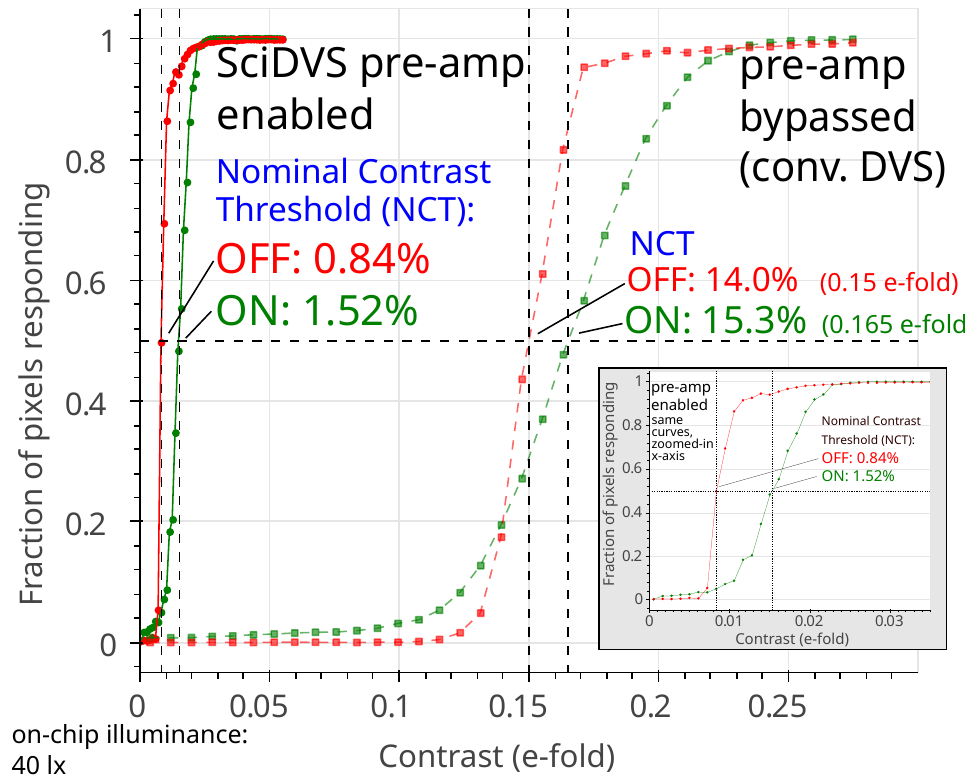}
\caption{SciDVS chip \xx{nct} sensitivity measurement. Compares SciDVS to preamp-disabled conventional \xx{dvs} operation.}
\label{fig:s_curve}
\end{figure}

\begin{figure}[tb]
\centering
\includegraphics[width=.8\columnwidth]{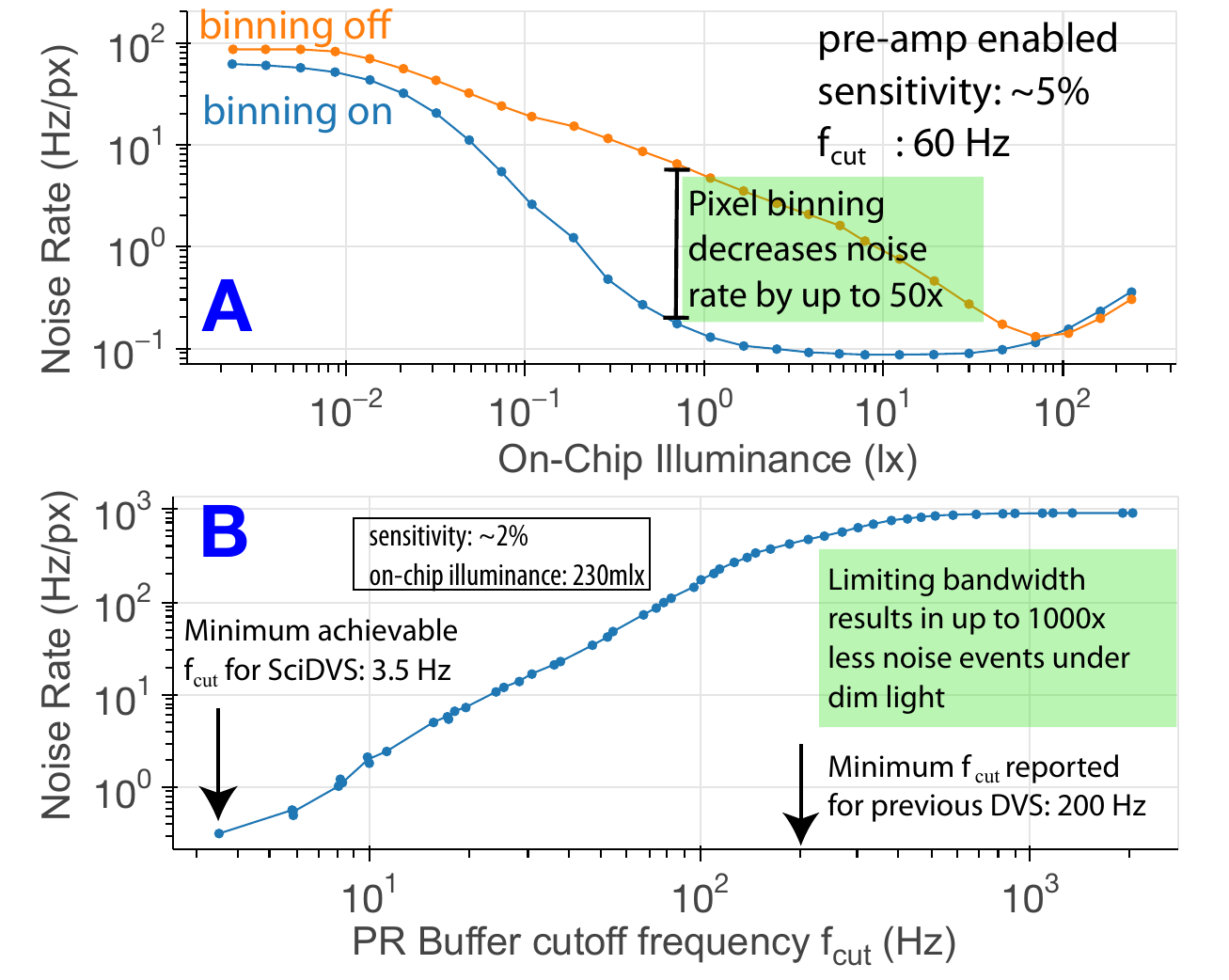}
\caption{Measured noise event rates. \textbf{A}: vs. illuminance with pixel binning on and off (array level). \textbf{B}: vs. PR buffer cutoff frequency $f_\text{cut}$ (test pixel)}
\label{fig:noise}
\end{figure}

\cref{fig:s_curve} shows the response of the SciDVS array to pulses of various contrast, from which the \xx{nct} is inferred. 
The curves show the fraction of pixels that respond to a step of a given contrast $C$, obtained by applying a high-contrast ($\approx$60\%) reset pulse, followed by a test step with contrast $C$. A pixel is considered to respond to a step if it outputs an event with the correct polarity in a \qty{200}{\milli\second} window following the step. The \xx{nct} is defined as the contrast step to which 50\% of the pixels respond.  
Using noisy settings underestimates the \xx{nct} due to constructive contribution of noise to the step (stochastic resonance)~\cite{mcreynolds2024reinterpretingthestepresponse}.
Therefore, the measurements were done at \qty{40}{\lux} chip illuminance with binning activated to minimize shot-noise (\cref{fig:noise}A) and thus make the most accurate estimate of \xx{nct}. 
With the preamp enabled, \xx{nct} OFF/ON of 0.84\%/1.52\% are achievable, with 0\% of the pixels responding with events of either polarity in response to a step of 0 contrast (i.e., no step) (see inset). This \xx{nct} is $>$10x smaller than standard \xx{dvs} mode (preamp bypassed).

\begin{figure}[tb]
\centering
\includegraphics[width=\columnwidth]{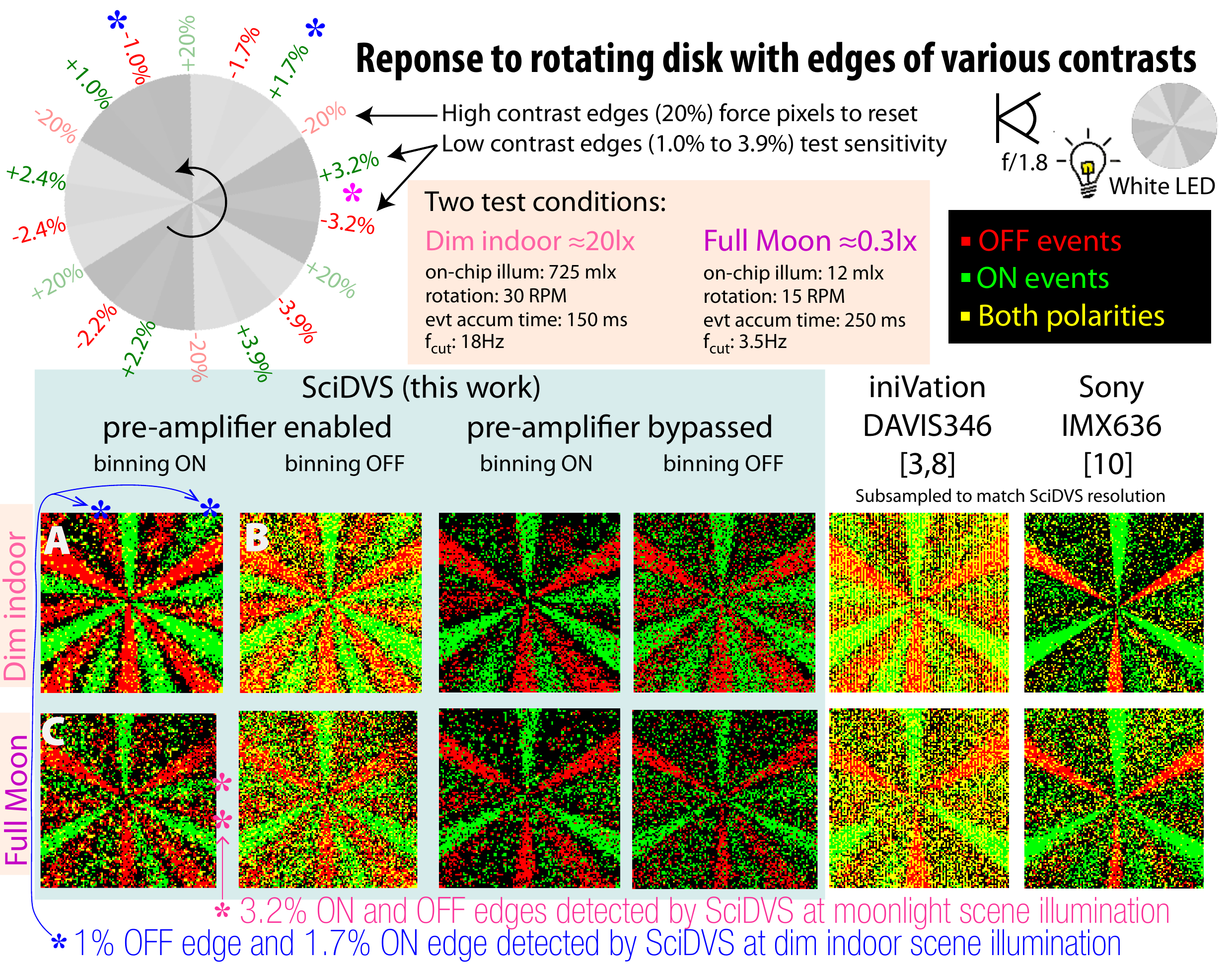}
\caption{SciDVS and commercial \xx{dvs} response to rotating disk with low contrast edges. On-chip illumination measured from calibrated test pixel photoreceptor output voltage. Video: \url{https://youtu.be/k6Cwx8yGOuU}.}
\label{fig:rotating_chart}
\end{figure}

\cref{fig:rotating_chart} compares \xx{dvs} sensitivity under low (\qty{0.7}{\lux}) and extremely low (\qty{12}{\milli\lux}) white LED chip illumination with two commercial \xx{dvs} cameras.
These chip illumination levels
correspond to scene illumination at dim indoor (\qty{20}{\lux}) and full moon (\qty{0.3}{\lux}) conditions~\cite{Delbruck1997-notes-practical-photometry}.
For both illumination conditions and all sensors, we optimized \xx{nct}, $\fcut$ and refractory period to maximize edge detection, while keeping the noise event rate under \qty{6}{\hertz/px}.
The rotating chart has large contrast 20\% edges to reset the pixels, and low contrast ON and an OFF test edges from the set of low contrasts 1\% to 3.9\%. 
Each image shows events accumulated over a \qty{200}{\milli\second} window.
To quantify sensitivity, an edge is considered detected if at least 50\% of the pixels respond with the correct polarity event during the \qty{200}{\milli\second} accumulation time.

Under dim indoor chart illumination
SciDVS reliably detects the OFF edge of 1\%, and the ON edge of 1.7\% (image \textbf{A} {\color{blue} blue *}). 
SciDVS clearly outperforms both SciDVS with preamp bypassed and the commercial \xx{dvs} where these edges are not visible in the images.
Turning SciDVS binning off (image \textbf{B}) increases edge detection probability through stochastic resonance, but the noise rate is much higher as can be seen from the large number of yellow pixels.

With full moon 
chart illumination,
SciDVS reliably detects the 3.2\% edges  (image \textbf{C}  {\color{magenta} magenta *}). These edges are not visible from SciDVS with the preamp disabled or from the commercial \xx{dvs}.

We can examine the plausibility of these results with  
\cref{eq:exposure}, which estimates the per-pixel photoelectrons $N$ transduced during the integration time $\tau\approx 1/(2\pi\fcut)$:
\begin{equation}
    \label{eq:exposure}
    N[\text{e}^-] \approx L[\si{\lux}] \times \qty{e4}{ph\per\lux\per\second\per\micro\meter\squared}\times \text{QE} \times (\qty{30}{\micro\meter})^2 \times 4 \times \tau[\text{s}] 
\end{equation}
\noindent where $L$ is the chip illuminance in \si{\lux}, the conversion factor of $10^4$ from \si{\lux} to visible photons comes from \cite{rose1973vision}, \xx{qe} $\approx 50\%$,
and the factor of 4 accounts for binning. 
The minimum possible \xx{dvs} photoreceptor noise predicted by \cite{graca2023optimalbiasing} 
is then $\sigma=\sqrt{2N}$ and $\sigma/N=\sqrt{2/N}$.

For $L$=\qty{0.7}{\lux} and $\fcut$=\qty{18}{\hertz}, 
$N$=\qty{110}{\kilo e^-} 
and $\sigma/N=0.4\%$, 
and for $L$=\qty{12}{\milli\lux} and $\fcut=\qty{3.5}{\hertz}$, 
N=\qty{10}{\kilo e^-} 
and $\sigma/N=1.4\%$. 
The observed sensitivities of 1.7\% and 3.2\% are respectively 4.2 and 2.2 times the predicted 1-$\sigma$ shot noise and therefore plausible.
Moreover, they suggest that SciDVS can be operated near the theoretical 2x shot noise limit.

\cref{fig:noise}A plots chip-level average pixel noise rate vs. illuminance with and without binning enabled for very sensitive \xx{nct}$\approx$5\%.
Binning reduces noise event rate per pixel in the \sfrac{1}{4} of active pixels by 50x at medium-low illumination.
This noise reduction effect of binning is also visible when comparing the SciDVS images in \cref{fig:rotating_chart}: The ones with binning are quieter;  the background wedges have fewer noise events.
\cref{fig:noise}B plots noise rate versus $\fcut$ with extremely low \xx{nct}$\approx$2\% and \qty{210}{\milli\lux} chip illuminance. Reducing $\fcut$ to \qty{3.5}{\hertz} reduces the noise event rate to \qty{0.1}{\hertz/px},
a factor of 1000x smaller than at the \qty{200}{\hertz} minimum $\fcut$ from the DAVIS346 \cite{graca2023optimalbiasing}. 

\begin{figure}[!t]
\centering
\includegraphics[width=.9\columnwidth]{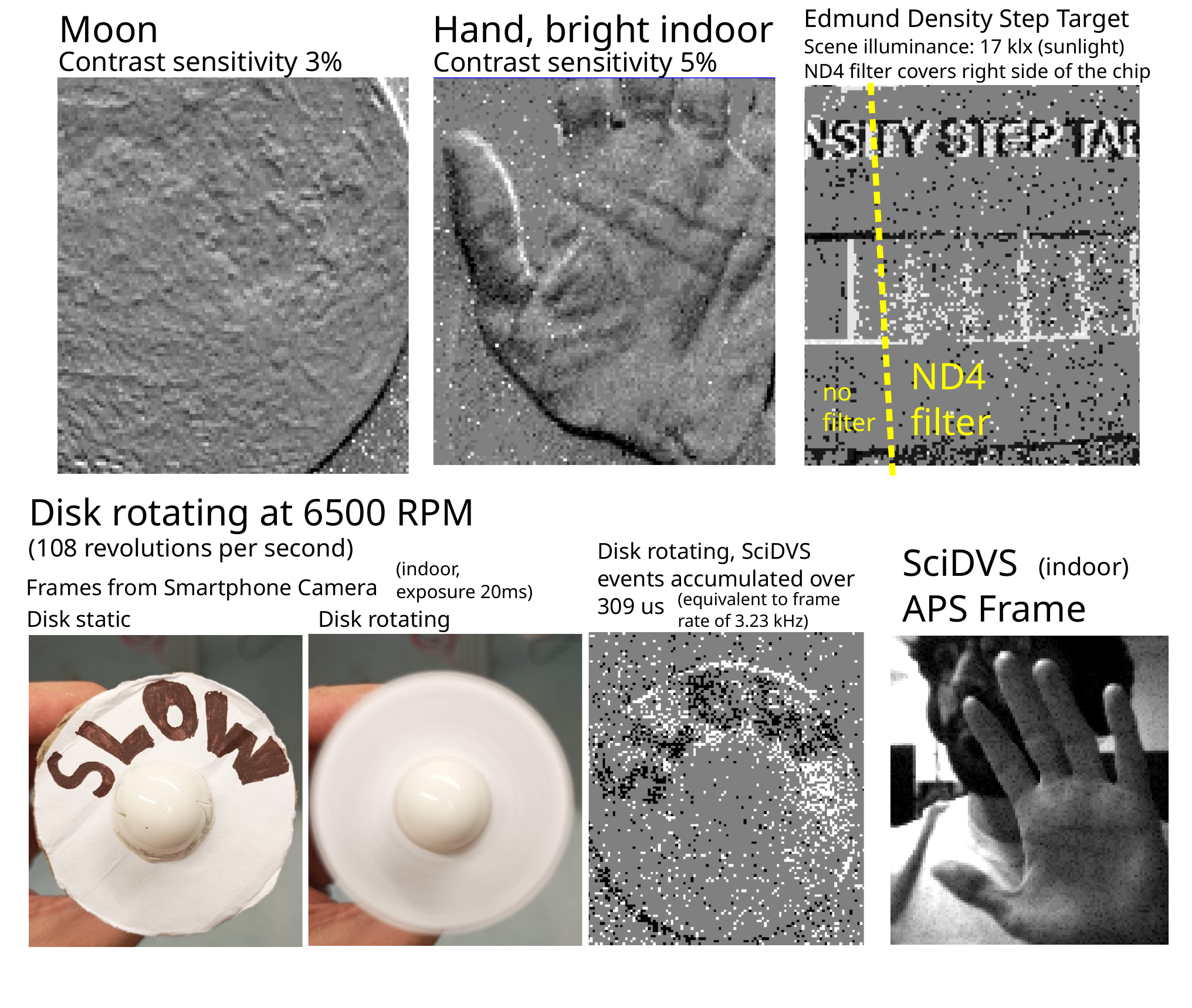}
\caption{SciDVS sample recordings. Video: \url{https://youtu.be/I3bse3z_ET0}.}
\label{fig:sample_recordings}
\end{figure}

\cref{fig:sample_recordings} shows several high sensitivity event accumulation frames, a high dynamic range scene, where part of the sensor was covered by an ND4 filter, and a SciDVS APS image.
High speed operation is demonstrated by accumulating events for \qty{309}{\micro\second} to reconstruct the ``SLOW'' text at an equivalent of  $\approx$\qty{3}{\kilo\hertz} frame rate. 

\cref{fig:comparison_table} is a comparison table, labeled die photo, and photo of the prototype camera.
SciDVS achieves 6x better sensitivity than commercial \xx{dvs} at 14x dimmer light~\cite{finateu2020backilluminated}, and achieves 1.7\% sensitivity at 3\,500x lower illumination than the previously reported sensitive prototype~\cite{yang2015dynamicvision}. 

\begin{table}[!t]
\centering
\caption{Comparison table, die photo with 2x2 pixel layout and camera.}
\includegraphics[width=\columnwidth]{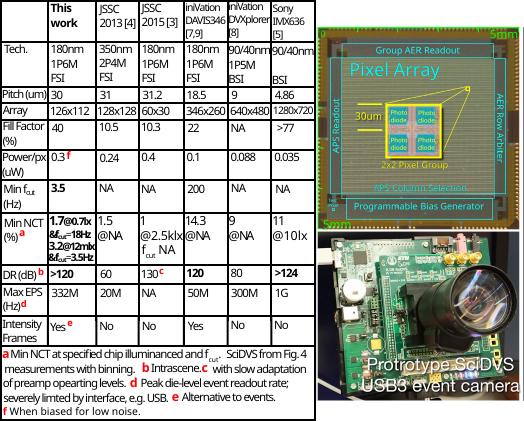}
\label{fig:comparison_table}
\end{table}

\section*{Acknowledgements:}
Funded by Swiss National Science Foundation project SCIDVS (200021\_185069). We thank SC Liu for their helpful comments.

\renewcommand*{\bibfont}{\footnotesize}
\footnotesize{\printbibliography}

\end{document}

%% file: acronyms.tex
\usepackage{xcolor} 

\usepackage[draft]{pdfcomment} 
\usepackage[hyperfirst=false,acronym,sort=none,shortcuts,nopostdot,style=super,nonumberlist,toc,nogroupskip]{glossaries}

\glsdisablehyper 

\definecolor{firstaccolor}{rgb}{0.2, 0.00, 0.00}

\newcommand{\accolor}[1]{\textcolor{Black}{#1}}

\newacronymstyle{myacro}
{%
  \GlsUseAcrEntryDispStyle{long-short}%
}%
{%
  \GlsUseAcrStyleDefs{long-short}%
}

\setacronymstyle{myacro}

\let\xx\tip

\newacronym{bpd}{BPD}{Buried Photodiode}
\newacronym{dvs}{DVS}{Dynamic Vision Sensor}
\newacronym{tc}{TC}{Temporal Contrast}
\newacronym{sf}{SF}{Source-Follower}
\newacronym{theta_on}{$\Theta_{\text{ON}}$}{ON threshold}
\newacronym{theta_off}{$\Theta_{\text{OFF}}$}{OFF threshold}
\newacronym{qe}{QE}{Quantum Efficiency}
\newacronym[description={Sigma/Mean}]{cov}{COV}{Coefficient Of Variation}

\newacronym[description={False Negative Rate; signal that is incorrectly classified as noise}]{fnr}{FNR}{False Negative Rate}
\newacronym[description={False Negative; signal event that is incorrectly classified as noise event}]{fn}{FN}{False Negative}
\newacronym[description={False Positive Rate; noise that is incorrectly classified as signal}]{fpr}{FPR}{False Positive Rate}
\newacronym[description={False Positive; noise event that is incorrectly classified as signal event}]{fp}{FP}{False Positive}
\newacronym[description={Guided Event Filtering: Joint Filtering of Intensity Images and Neuromorphic Events}]{gef}{GEF}{Guided Event Flow}
\newacronym[description={Inter Spike Interval (nomenclature from neuroscience)}]{isi}{ISI}{Inter Spike Interval}
\newacronym[description={MAC (Multiply-Accumulate) is the basic operation of signal processing and artificial neural networks. One MAC is 2 Op.}]{mac}{MAC}{Multiply-Accumulate}
\newacronym[description={Multipurpose block random access memory module in FPGA}]{bram}{BRAM}{Block RAM}
\newacronym[description={Register Transfer Logic intermediate form, consisting of combinational and synchronous register logic cells}]{rtl}{RTL}{Register Transfer Logic}
\newacronym[description={Single Threshold Metric; a measure of the ROC TPR/FPR tradeoff at one discrimination threshold}]{stm}{STM}{Single Threshold Metric}
\newacronym[description={Surface of Active Event; image of latest event timestamps, same as Timestamp Image}]{sae}{SAE}{Surface of Active Events}
\newacronym[description={System on Chip; FPGA with embedded programmable processor}]{soc}{SoC}{System on Chip}
\newacronym[description={Time Surface; image of age of events relative to a particular event}]{ts}{TS}{Time Surface}
\newacronym[description={Timestamp Image; 2D image of latest event timstamps, similar to Surface of Active Events}]{ti}{TI}{Timestamp Image}
\newacronym[description={Timestamp+Polarity Image; 2D image of latest event timstamps and +/- brightness change polarites}]{tpi}{TPI}{Timestamp+Polarity Image}
\newacronym[description={True Negative Rate; noise that is correctly classified as noise}]{tnr}{TNR}{True Negative Rate}
\newacronym[description={True Negative; noise that is correctly classified as noise}]{tn}{TN}{True Negative}
\newacronym[description={True Positive Rate; signal that is correctly classified as signal}]{tpr}{TPR}{True Positive Rate}
\newacronym[description={True Positive; signal event that is correctly classified as signal}]{tp}{TP}{True Positive}
\newacronym[longplural={Convolutional Neural Networks}]{cnn}{CNN}{Convolutional Neural Network}
\newacronym[longplural={First In First Out memories}]{fifo}{FIFO}{First In First Out memory}
\newacronym{adc}{ADC}{Analog to Digital Converter}
\newacronym{aer}{AER}{Address Event Protocol}
\newacronym{aps}{APS}{Active Pixel Sensor}
\newacronym{asic}{ASIC}{Application Specific Integrated Circuit}
\newacronym{auc}{AUC}{Area Under the Curve}
\newacronym{baf}{BAF}{Background Activity Filter}
\newacronym{ba}{BA}{Background Activity}
\newacronym{bmof}{BMOF}{Block Matching Optical Flow}
\newacronym{bm}{BM}{Block Matching}
\newacronym{bp}{BP}{Back Propagation}
\newacronym{cfa}{CFA}{Color Filter Array}
\newacronym{cf}{CF}{Complementary Filter}
\newacronym{cg}{CG}{Convolutional Gated Recurrent Unit Network}
\newacronym{cis}{CIS}{CMOS Image Sensor}
\newacronym{cmae}{CMAE}{Cross-Modality Attention Enhancement}
\newacronym{contrastmaximization}{CM}{Contrast Maximization}
\newacronym{cots}{COTS}{Commodity Off-The-Shelf}
\newacronym{cpu}{CPU}{Central Processing Unit}
\newacronym{cv}{CV}{Computer Vision}
\newacronym{davis}{DAVIS}{Dynamic and Active pixel Vision Sensor}
\newacronym{dba}{DBA}{Dynamic Background Activity noise filtering algorithm}
\newacronym{dnn}{DNN}{Deep Neural Network}
\newacronym{dof}{DOF}{Degree of Freedom}
\newacronym{dolp}{DoLP}{Degree of Linear Polarization}
\newacronym{dram}{DRAM}{Dynamic RAM}
\newacronym{drcn}{DRCN}{Deep Recurrent Convolutional Network}
\newacronym{dr}{DR}{Dynamic Range}
\newacronym{dsp}{DSP}{Digital Signal Processing unit}
\newacronym{dwf}{DWF}{Double Window Filter}
\newacronym{e2pd}{E2PD}{Events to Polarization Dataset}
\newacronym{e2p}{E2P}{Events to Polarization}
\newacronym{edflow}{EDFLOW}{Event-driven Optical Flow}
\newacronym{edncnn}{EDnCNN}{Event Denoising CNN}
\newacronym{edp}{EDP}{Event Denoising Precision}
\newacronym{efast}{EFAST}{Event-Based time surface FAST}
\newacronym{epm}{EPM}{Event Probability Mask}
\newacronym{fast}{FAST}{Features from Accelerated Segment Test}
\newacronym{feast}{FEAST}{Feature Extraction with Adaptive Selection Thresholds }
\newacronym{flipflop}{FF}{Flip-Flop}
\newacronym{fom}{FOM}{Figure of Merit}
\newacronym{fpga}{FPGA}{Field Programmable Gate Array}
\newacronym{fpn}{FPN}{Fixed Pattern Noise}
\newacronym{fps}{FPS}{Frames Per Second}
\newacronym{fsae}{FSAE}{Filtered Surface of Active Events}
\newacronym{fwf}{FWF}{Fixed Window Filter}
\newacronym{gpu}{GPU}{Graphics Processing Unit}
\newacronym{gt}{GT}{Ground Truth}
\newacronym{hdl}{HDL}{Hardware Description Language}
\newacronym{hdr}{HDR}{High Dynamic Range}
\newacronym{hls}{HLS}{High Level Synthesis}
\newacronym{icm}{ICM}{Iterated Conditional Modes}
\newacronym{id}{ID}{Index Decay}
\newacronym{iir}{IIR}{Infinite Impulse Response}
\newacronym{imu}{IMU}{Inertial Measurement Unit}
\newacronym{inceptiveevent}{IE}{Inceptive Event}
\newacronym{iot}{IoT}{Internet of Things}
\newacronym{ip}{IP}{Intellectual Property}
\newacronym{its}{ITS}{Invariant Time Surface}
\newacronym{knn}{KNN}{$K$-Nearest-Neighbor clustering}
\newacronym{li}{LI}{Leaky Integrator}
\newacronym{lk}{LK}{Lucas-Kanade}
\newacronym{lpips}{LPIPS}{Learned Perceptual Image Patch Similarity}
\newacronym{lut}{LUT}{LookUp Table}
\newacronym{mlpf}{MLPF}{MultiLayer Perceptron denoising Filter}
\newacronym{mlp}{MLP}{Multilayer Perceptron}
\newacronym{ml}{ML}{Machine Learning}
\newacronym{mpeg}{MPEG}{Motion Picture Experts Group}
\newacronym{mse}{MSE}{Mean Squared Error}
\newacronym{na}{NA}{Numerical Aperture}
\newacronym{nnb}{NNb}{Nearest Neighbor}
\newacronym{of}{OF}{Optical Flow}
\newacronym{onf}{ONF}{Order(N) Filter}
\newacronym{pcb}{PCB}{Printed Circuit Board}
\newacronym{pdavis}{PDAVIS}{Polarization Dynamic and Active pixel VIsion Sensor}
\newacronym{pd}{PD}{photodiode}
\newacronym{pe}{PE}{Processing Element}
\newacronym{pfa}{PFA}{Polarization Filter Array}
\newacronym{pl}{PL}{programmable Logic}
\newacronym{por}{POR}{Positive Output Ratio}
\newacronym{prm}{PRM}{Pixel Rendering Module}
\newacronym{ps}{PS}{Processing System}
\newacronym{pugm}{PUGM}{Probabilistic Undirected Graph Model}
\newacronym{qwp}{QWP}{Quarter Wave Plate}
\newacronym{ram}{RAM}{Random Access Memory}
\newacronym{ransac}{RANSAC}{Random Sample and Consensus}
\newacronym{ratp}{RATP}{Recursive Adaptive Temporal Pooling}
\newacronym{rb}{RB}{Residual Block}
\newacronym{relu}{ReLU}{Rectified Linear Unit}
\newacronym{roc}{ROC}{Receiver Operating Characteristic}
\newacronym{roi}{ROI}{Region of Interest}
\newacronym{rpmd}{RPMD}{Relative Plausibility Measure of Denoising}
\newacronym{rpm}{RPM}{Revolutions per Minute}
\newacronym{rppp}{RPPP}{Rich Polarization Pattern Perception}
\newacronym{sad}{SAD}{Sum of Absolute Differences}
\newacronym{sm}{SM}{Supplementary Material}
\newacronym{snr}{SNR}{Signal to Noise Ratio}
\newacronym{soa}{SOA}{state of the art}
\newacronym{sram}{SRAM}{Static RAM}
\newacronym{stcf}{STCF}{SpatioTemporal Correlation Filter}
\newacronym{tda}{TDA}{Time Decay Adapted}
\newacronym{td}{TD}{Time Decay}
\newacronym{timsl}{TS}{time slice}
\newacronym{usb}{USB}{Universal Serial Bus}
\newacronym{vga}{VGA}{Video Graphics Adaptor}
\newacronym{vhdl}{VHDL}{Very High-Speed Integrated Circuit Hardware Description Language}
\newacronym{zoh}{ZOH}{Zero-Order Hold}
\newacronym{nct}{NCT}{Nominal Contrast Threshold}
